\documentclass[11pt,twoside]{article}
\usepackage{asp2006}
\usepackage{graphics}
\usepackage{graphicx}
\usepackage{lscape}
\def\edcomment#1{\iffalse\marginpar{\raggedright\sl#1\/}\else\relax\fi}
\reversemarginpar

\begin{document}
\title{the Cosmically Depressed:\\ Life, Sociology and Identity of Voids}
\author{R. van de Weygaert, E. Platen, E. Tigrak, J. Hidding,\\ J.M. van der Hulst}
\affil{Kapteyn Astronomical Institute, University of Groningen, 9700 AV  Groningen, the Netherlands}
\author{M.~A.~Arag\'on-Calvo}
\affil{Dept. Physics \& Astronomy, The Johns Hopkins University, 3701 San Martin Drive, Baltimore, MD 21218, USA}
\author{K. Stanonik, J.H. van Gorkom}
\affil{Department of Astronomy, Columbia University, Mail Code 5246, 550 West 120th Street, New York, NY 10027, USA}

\begin{abstract}
In this contribution we review and discuss several aspects of Cosmic Voids, 
as a background for our Void Galaxy Survey (accompanying paper by Stanonik et al.). 
Following a sketch of the general characteristics of void 
formation and evolution, we describe the influence of the environment on 
their development and structure and the characteristic hierarchical buildup of 
the cosmic void population. In order to be able to study the resulting 
tenuous void substructure and the galaxies populating the interior of 
voids, we subsequently set out to describe our parameter free tessellation-based 
watershed void finding technique. It allows us to trace the outline, shape 
and size of voids in galaxy redshift surveys. The application of this 
technique enables us to find galaxies in the deepest troughs of the cosmic 
galaxy distribution, and has formed the basis of our void galaxy program.
\end{abstract}

\vspace{-0.5cm}
\section{Cosmic Depressions}
{\it Cosmic Voids} have been known as a feature of galaxy surveys since the first surveys were compiled \citep{chincar1975,gregthomp1978,einasto1980}.  
They are enormous regions with sizes in the range of $20-50h^{-1}$ Mpc that are practically devoid of any galaxy, usually roundish 
in shape and occupying the major share of space in the Universe. Forming a key component of the {\it Cosmic Web} \citep{bondweb1996}, 
they are surrounded by elongated filaments, sheetlike walls and dense compact clusters.  Following the discovery by ~\cite{kirshner1981} 
of the most dramatic specimen, the Bo\"otes void, a hint of their central position within a weblike arrangement 
came with the first CfA redshift slice \citep{lapparent1986}. This view has been dramatically endorsed and expanded by the redshift maps of 
the 2dFGRS and SDSS surveys \citep{colless2003,tegmark2004}. They have established voids as an integral component of the 
Cosmic Web. The 2dFGRS maps and SDSS maps are telling illustrations of the ubiquity 
and prominence of voids in the cosmic galaxy distribution. There are a variety of reasons why voids are cosmologically 
interesting. They are a prominent aspect of the Megaparsec Universe, instrumental in the spatial 
organization of the Cosmic Web. They also contain a considerable amount of information on the underlying 
cosmological scenario and on global cosmological parameters \citep[e.g.][]{rydmel1996,parklee2007,lavaux2009}. A final 
important aspect is that the pristine low-density environment of voids 
represents an ideal and pure setting for the study of galaxy formation and the influence of cosmic environment 
on the formation of galaxies \citep[e.g.][]{peebles2001}. 

\section{Void Life: Formation and Evolution}
\label{sec:voidevol}
At any cosmic epoch the voids that dominate the spatial matter distribution are a manifestation of the cosmic structure 
formation process reaching a non-linear stage of evolution.

Voids emerge out of the density troughs in the primordial Gaussian field of density fluctuations. Early theoretical models 
of void formation concentrated on the evolution of isolated voids \citep{hoffshah1982,icke1984,edbert1985,blumenth1992}. 
As a result of their underdensity voids represent a region of weaker gravity, resulting in an effective repulsive peculiar 
gravitational influence. Initially underdense regions therefore expand faster than the Hubble flow, and thus expand 
with respect to the background Universe. As voids expand, matter is squeezed in between them, and sheets and filaments form 
the void boundaries. This view is supported by numerical studies and computer simulations of the gravitational evolution of 
voids in more complex and realistic configurations \citep{weykamp1993,colberg2005}. 

\section{Void Sociology}
\label{sec:voidsocio}
\begin{figure}[h]
\begin{center}
\mbox{\hskip -0.0truecm\includegraphics[width=1.00\textwidth]{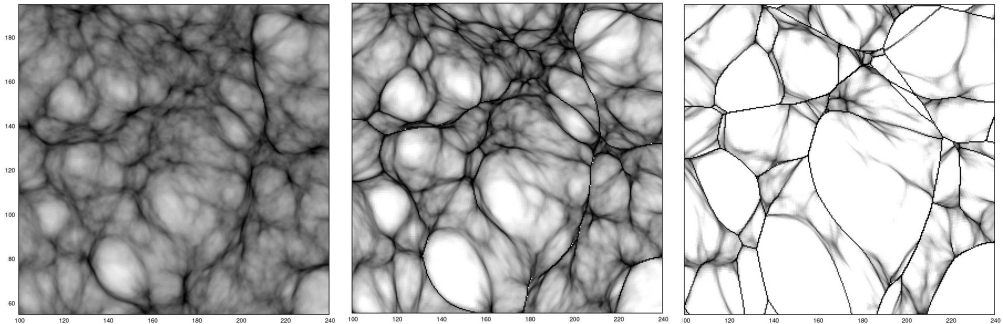}}
\end{center}
\begin{center}
\vskip -0.5truecm
\caption{The evolution of cosmic structure according to the adhesion approximation (see text), 
in a 2-D (Eulerian) region of size 120h$^{-1}$Mpc, at three different epochs. From left to right: 
a=0.15,0.30,0.70.}
\end{center}
\vskip -0.75truecm
\label{fig:adhesionhier}
\end{figure}

Computer simulations of the gravitational evolution of voids in realistic cosmological environments do 
show a considerably more complex situation than that described by idealized spherical or ellipsoidal models 
\citep{weykamp1993,colberg2005}. In recent years the huge increase in computational resources has 
enabled {\rm N}-body simulations to resolve in detail the intricate substructure of voids within the context 
of hierarchical cosmological structure formation scenarios. They confirm the 
theoretical expectation of voids having a rich substructure as a result of their hierarchical buildup 
(see e.g. fig.~\ref{fig:adhesionhier}). 

It leads to a considerably modified view of the evolution of voids. One aspect concerns the dominant 
environmental influence on the evolution of voids. To a large extent the shape and mutual alignment of 
voids is dictated by the surrounding large scale structure and by large scale gravitational tidal 
influences. Equally important is the role of substructure within the interior of voids. This, 
and the interaction with the surroundings, turn out to be essential aspects of the {\it hierarchical} 
evolution of the void population in the Universe. 

\subsection{Environmentally Shaped, Tidally Directed}
\label{sec:environ}
\cite{icke1984} pointed out that any (isolated) aspherical underdensity will become more spherical as it expands. 
In reality, voids will never reach sphericity. Even though voids tend to be less flattened or elongated than 
the halos in the dark matter distribution, they are nevertheless quite nonspherical. The flattening is a result of 
large scale dynamical and environmental factors \citep{platen2008}. Even while their internal dynamics pushes them to a more 
spherical shape they will never be able to reach perfect sphericity before encountering surrounding structures such 
as neighbouring voids, overdense filaments and planar walls. Even more important may be the fact that, for voids, 
external tidal influences remain important: voids will always be rather moderate densities perturbations since they can 
never be more underdense than $\delta=-1$. External tidal forces are responsible for a significant 
anisotropic effect in the development of the voids. In extreme cases they may even cause complete collapse 
of the void.   

\subsection{Void Hierarchy: to Merge or to Collapse}
In some sense voids have a considerably more complex evolutionary path than overdense halos. Many small primordial 
density troughs may exist within a larger underdense region. Their evolution, fate and interaction are key towards 
understanding the hierarchical buildup of the void network. Two processes dictate the evolution of voids: their 
{\it merging} into ever larger voids as well as the {\it collapse} and disappearance of small ones embedded in 
overdense regions. By identifying and assigning critical density values to the two evolutionary void processes of 
merging and collapse, \cite{shethwey2004} managed to describe the hierarchical evolution of the void population in 
terms of a two-barrier {\it excursion set} formulation \citep{bond1991}, akin to the evolution of overdense halos. 
An important guideline to this was the heuristic void model simulations by \cite{dubinski1993}. A 
significant contribution for a proper theoretical insight into the unfolding void hierarchy is the theoretical void 
study by \cite{sahni1994} within the context of a Lagrangian adhesion model description. 

An impression of this complex lifeline may be obtained from the sequence of timesteps depicted in fig.~\ref{fig:adhesionhier}. 
The figure shows the emerging weblike structure in a $\Lambda$CDM universe, in Eulerian space, following the adhesion 
approximation. One may see how the initially intricate weblike network in 
the interior of the large central underdense region 
gradually disappears as voids merge while the internal boundaries gradually fade away. In particular near the 
boundaries of large voids we may see the second void process, that of collapse of voids. It manifests itself 
in the form of a shearing and squeezing of less prominent voids, as a result of the expansion of prominent 
neighbouring voids or of the tidally induced filamentary or planar collapse of the weblike mass concentrations 
at edges of the voids. 

The merging of subvoids within a large void's interior usually follows the emergence of these small-scale 
depressions as true voids. Once they do, their expansion tends to slow down. When the adjacent subvoids meet up, 
the matter in between is squeezed in thin walls and filaments. The peculiar velocities perpendicular to the 
void walls are mostly suppressed, resulting in a primarily tangential flow of matter within their mutual 
boundaries and the gradual fading of these structures while matter evacuates along the walls and filaments 
towards the enclosing boundary of the emerging void \citep{dubinski1993}. The final result is the merging and 
absorption of the subvoids in the larger void which gradually emerges from their embedding underdensity. As far 
as the void population is concerned only the large void counts, while the faint and gradually fading imprint of 
the original outline of the subvoids remains as a reminder of the initial internal substructure. 

The second void process, that of the collapse of mostly small and medium-sized voids, is responsible for 
the radical dissimilarity between void and halo populations. If a small scale minimum is embedded in a sufficiently 
high large scale density maximum, then the collapse of the larger surrounding region will eventually squeeze the 
underdense region it surrounds: small-scale voids will vanish when the region around them fully collapses. 
The manifest anisotropic shearing of collapsing voids near the boundaries of prominent voids (fig.~\ref{fig:adhesionhier}) 
is an indication for the important role of tidal forces in bringing about their demise. 

\subsection{Void Infrastructure}
An important issue within the hierarchically proceeding evolution of voids and the Cosmic Web 
is the fate of its substructure. In voids the diluted and diminished infrastructure remains 
visible, at ever decreasing density contrast, as cinders of the earlier phases of the {\it void 
hierarchy} in which the substructure stood out more prominent

{\rm N}-body simulations show that voids do retain a rich infrastructure. Examples such as the images of 
the Millennium simulation \citep{springmillen2005} show that while void substructure does fade, it does not 
disappear. We may find structures ranging from filamentary and sheetlike structures to a population of low 
mass dark matter halos and galaxies. Although challenging, these may also be seen in the observational 
reality. The galaxies that populate the voids do currently attract quite some attention (see next section). 
Also, the SDSS galaxy survey has uncovered a substantial level of 
substructure within the Bo\"otes void.

\section{In search of Voids}
In order to be able to test theoretical predictions and compare them with the voids in observed reality, or 
those in the complex environment of N-body simulations, we need a strict and proper definition of a void. 
Voids are distinctive and striking features of the cosmic web, yet identifying them and tracing their outline 
within the complex geometry of the galaxy and mass distribution in galaxy surveys and simulations has proven 
to be a nontrivial issue. The watershed-based WVF 
algorithm of \cite{platen2007} aims to avoid issues of both sampling density and shape. The WVF is defined with respect 
to the DTFE density field, assuring optimal sensitivity to the morphology of spatial structures and an unbiased probe 
over the full range of substructure in the mass distribution.

\begin{figure}[h]
\begin{center}
\vskip -0.0truecm
\mbox{\hskip -0.18truecm\includegraphics[width=1.00\textwidth]{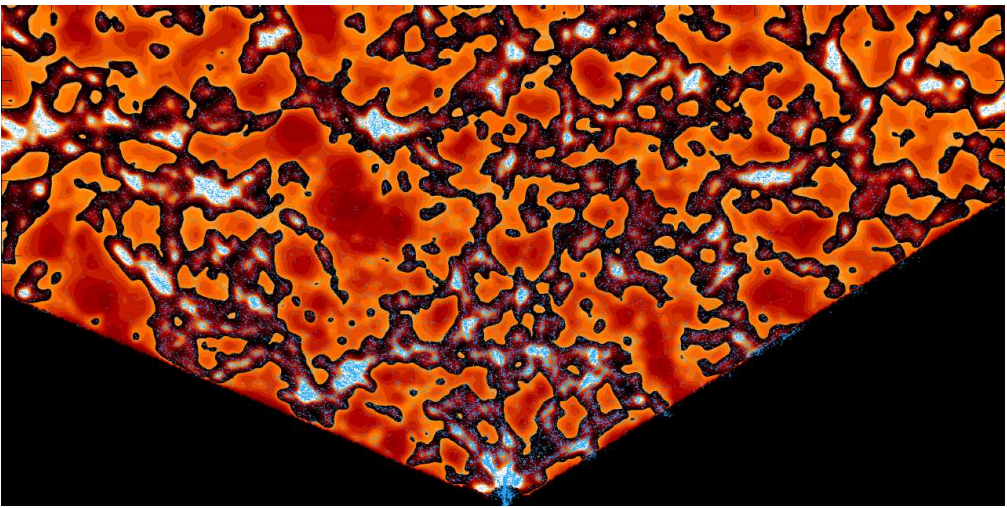}}
\end{center}
\begin{center}
\vskip -0.5truecm
\caption{A visualization of the DTFE SDSS density field in a 12h$^{-1}$ Mpc wide slice. The contour levels 
of the 1h$^{-1}$ Mpc smoothed field are roughly divided between underdense and overdense regions. Galaxies 
are indicated by blue dots. Some of the most prominent features have been indicated: the \textit{Bo\"otes 
Supervoid} and the large supervoid identified by Bahcall \& Soneira (1982). Also the large overdense 
\textit{(Coma) Great Wall} and the \textit{Hercules Supercluster} within this wall are indicated. From 
Platen 2009.}
\end{center}
\vskip -0.75truecm
\label{fig:sdssdtfe}
\end{figure}

\subsection{Tiling for Mapping: DTFE} 
When studying the cosmic matter distribution in the nearby Universe, the main source of information is that 
of the spatial location of galaxies. Also computer simulations of cosmic structure formation follow the 
evolving matter distribution by means of a large sample of discrete tracers, N-body particles. For the purpose 
of analyzing the topological intricacies of the cosmic web, and for tracibng the outline of voids and filaments, we
have developed a technique based on a natural adaptive tiling of space defined by the sampling datapoints, 
be it galaxies or N-body particles. 

The DTFE procedure, the Delaunay Tessellation Field Estimator \citep{schaapwey2000,schaap2007,
weyschaap2009}, is based on the Voronoi and its dual Delaunay tessellation \citep{okabe2000} in the 
survey/simulation volume. The method exploits the adaptivity of these tessellation cells to the local density and geometry 
of the generating spatial point process. The volume of the (contiguous) Voronoi cell around a sample galaxy is used as 
a zeroth order measure for the density at the location of the galaxy. The obtained density values are subsequently 
interpolated throughout the survey volume, using the Delaunay triangulation as the adaptive and irregular grid for 
a linear interpolation procedure. 

In addition to the computational efficiency of the procedure, the density maps produced by DTFE have the virtue of retaining 
the anisotropic and hierarchical structures which are so characteristic of the Cosmic Web \citep{schaap2007,weyschaap2009}. 
The recent in-depth analysis by \cite{platen2009} has shown that for very large point samples, DTFE even outperforms 
more elaborate high-order methods with respect to quantitative and statistical evaluations of the density field. As a result, 
the DTFE density field can be used for objectively tracing structural features such as walls, filaments and voids. 
This is amply illustrated by the detailed map shown in fig.~\ref{fig:sdssdtfe}, showing the distribution of galaxies 
in a slice through the SDSS galaxy redshift survey. 

\begin{figure}[t]
\begin{center}
\vskip -0.25truecm
\mbox{\hskip -0.18truecm\includegraphics[width=0.95\textwidth]{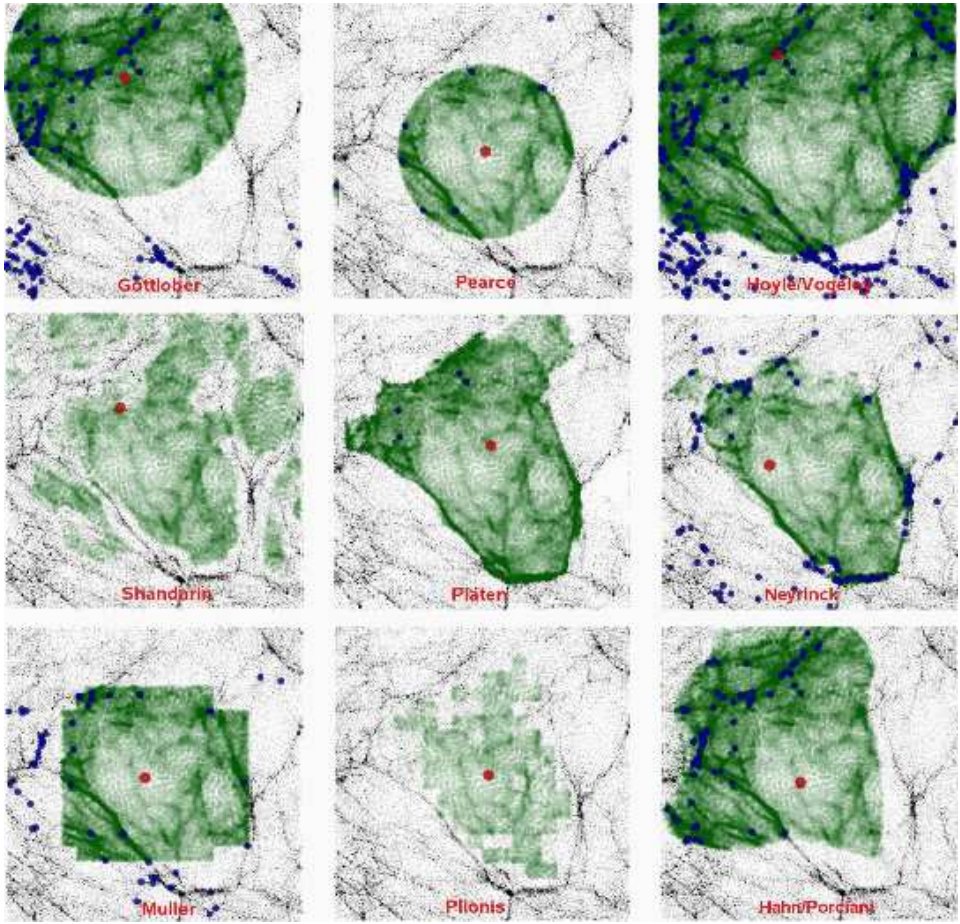}}
\end{center}
\begin{center}
\vskip -0.5truecm
\caption{A compilation of void finders. The 9 frames illustrate the performance of different void 
finders with respect to a central voids in the milli-Millennium simulation. The N-body dark matter particles 
are depicted as black points. The blue dots locate (semi-analytically modelled) galaxies within the 
central void region. For each void finder, the identified void region is coloured green with the void 
centre marked by a red point. From Colberg et al. 2008.}
\end{center}
\vskip -1.0truecm
\label{fig:voidfind}
\end{figure}

\subsection{A Watershed Search}
When studying the topological and morphological structure of the cosmic matter distribution 
in the Cosmic Web, it is convenient to draw the analogy with a landscape. \textit{Valleys} represent the 
large underdense voids that define the cells of the Cosmic Web. Their 
boundaries are \textit{sheets} and \textit{ridges}, defining the network 
of walls, filaments and clusters that defines the Cosmic Web.

A commonly used method in Image Analysis is the Watershed Transform (WST). It is a concept defined within the 
context of mathematical morphology, and was first introduced by \cite{beulan1979}. It is widely used for segmenting 
images into distinct patches and features. The basic idea behind the WST stems from geophysics, where it is used to 
delineate the boundaries of separate domains, i.e. {\it basins} into which yields of e.g. rainfall will collect. 

The word {\it watershed} finds its origin in the analogy of the procedure with that of 
a landscape being flooded by a rising level of water. Suppose we have a surface in the shape of a landscape, 
which is pierced at the location of each of the minima. As the water-level rises a growing fraction 
of the landscape will be flooded by the water in the expanding basins. Ultimately basins will meet 
at the ridges defined by {\it saddle-points} and {\it maxima} in the density field. The final 
result of the completely immersed landscape is a division of the landscape into individual 
cells, separated by the {\it ridge dams}.

The watershed transform was first introduced in a cosmological context as an objective technique 
to identify and outline voids in the cosmic matter and galaxy distribution \citep{platen2007,platen2009}. 
Following the density field-landscape analogy, the Watershed Void Finder (WVF) method identifies 
the underdense void patches in the cosmic matter distribution with the watershed basins. 

With respect to the other void finders the watershed algorithm has several advantages. The virtues of 
WVF may be appreciated from the comparison of its performance with a few other void finding algorithms 
in fig.~\ref{fig:voidfind} (from \cite{colberg2008}). The situation involves the region in and around a 
large void in the milli-Millennium simulation \citep{springmillen2005}. Because it identifies a void segment 
on the basis of the crests in a density field surrounding a density minimum it is able to trace the void 
boundary even though it has a distorted and twisted shape. Also, because the contours around well chosen 
minima are by definition closed the transform is not sensitive to local protrusions between two adjacent voids. 
The main advantage of the WVF is that for an ideally smoothed density field it is able to find voids in an entirely 
parameter free fashion (in case there is noi noise in the data; in less ideal, and realistic, circumstances 
a few parameters have to be set for filtering out discreteness noise). Because the watershed works 
directly on the topology of the field, and does not involve a predefined geometry/shape. This is the 
reason behind one its main advantages, its independence of assumptions on the shape and size of voids 
\footnote{WVF shares this virtue with a similar tessellation-based void finding method, 
ZOBOV (Neyrinck 2008).}. The quoted vritues make WVF particularly suited for the analysis of the 
hierarchical void distribution expected in the commonly accepted cosmological scenarios. 


\begin{figure}[h]
\vskip -0.25truecm
\begin{center}
\mbox{\hskip -0.0truecm\includegraphics[width=0.95\textwidth]{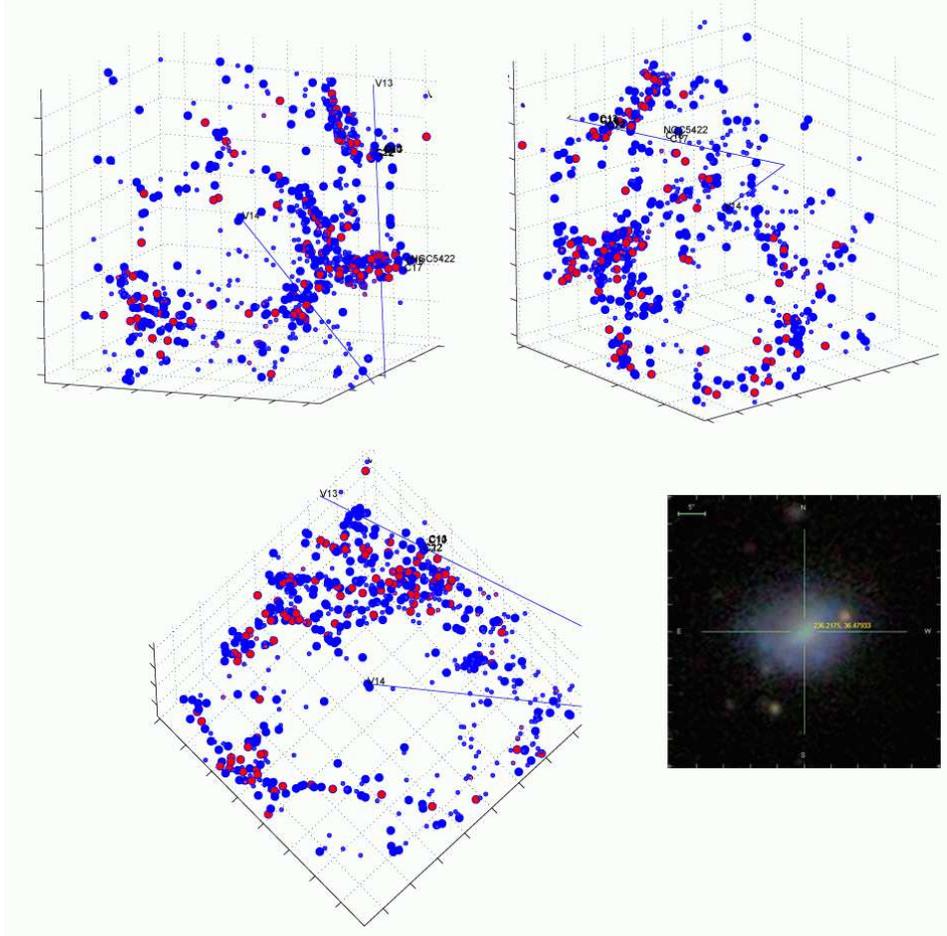}}
\end{center}
\vskip -1.5truecm
\begin{center}
\caption{The SDSS galaxy distribution in and around the void in which void galaxy voidp14 of our void 
galaxy sample is located. Bright galaxies have large dots (B$<$-16), fainter ones are depicted by 
smaller dots. Redder galaxies, with g-r$>$0.6 are indicated by red dots.}
\end{center}
\vskip -1.25truecm
\label{fig:voidgalsel}
\end{figure}

\section{Lonely Guards of the Cosmos: Void Galaxies}
\label{sec:galselect}
The pristine and isolated nature of the void environment represents an ideal setting for the study of 
galaxy formation. Largely unaffected by the complexities and processes modifying galaxies in high-density environments, 
the isolated void regions are expected to hold important clues to the formation and evolution of galaxies and our 
understanding of the environmental influences on galaxy formation. The presence and nature of void galaxies is also 
important for at least two additional reasons. The existence, or rather the apparent underabundance, of galaxies in 
the void regions may present a challenge for currently favoured galaxy formation theories \citep{peebles2001}. Interesting 
is also the extent to which galaxies trace substructure in voids. Such tenuous features would be fossil remnants of the 
hierarchical buildup of the Cosmic Web \citep{shethwey2004}. 

We have started a program to study the properties, the environment and their relationship of a complete sample of 60 void 
galaxies, probing a diversity in cosmic voids. The preliminary results of this campaign are described in the 
accompanying contribution by Stanonik et al. \citep[also see][]{stanonik2009a,stanonik2009b}. Our ability to outline 
voids in the SDSS galaxy distribution, by means of the WVF/DTFE procedure described in the previous section, has enabled 
us to identify a sample of galaxies which reside in the central and most underdense areas of cosmic voids. In summary, 
it allows us to find the emptiest regions and most lonely galaxies in the Local Universe. Upon having obtained the complete 
list of voids in the SDSS survey volume, and the void galaxies within their realm, we evaluate for each of the galaxies 
whether it conforms to a set of additional criteria. The galaxy should be 
\begin{itemize}
\item[$\bullet$] located in the interior central region of clearly defined voids, 
if possible near the centre, and be as far removed from the boundary of the voids as possible.  
\item[$\bullet$] removed from the edge of the SDSS survey volume, as we do not wish to have galaxies 
in voids which extend past the edge of the SDSS coverage. 
\item[$\bullet$] be isolated (not a member of a group)
\item[$\bullet$] not be within $\approx 750$km/s from a foreground or background cluster. This 
assured us that the presence in a void of a galaxy could not be attributed to finger of god  
of god effect. 
\item[$\bullet$] for observational purposes , we prefer galaxies to be within a 
redshift $0.01<z<0.02$. It allows sufficient sensitivity and resolution of the gas 
structure and kinematics in the galaxies. 
\end{itemize}
We also choose to select each sample galaxy to be in a different void. Note that there is no a priori 
selection on luminosity or colour of the void galaxies in our sample. The sample contains a representative 
number of early type galaxies.

A very nice example of the location of one of our void galaxies, nr. 14 in the pilot sample list, 
is shown in fig.~\ref{fig:voidgalsel}. It shows the SDSS galaxy distribution in and around the 
void in which the void galaxy (SDSS image included) is located. By showing the spatial galaxy 
distribution from three different perspectives, we seek to evoke a genuine 
three-dimensional impression. 

\acknowledgements We wish to acknowledge Ravi Sheth, Michael Vogeley, Jim Peebles and Bernard Jones for 
many useful discussions and key contributions to aspects of the discussed issues.

\end{document}